\documentclass{PoS}

\def\apj{\emph{ApJ.}}

\def\aap{\emph{A.\& A.}}
\def\apjs{\emph{ApJS}}

\title{VERITAS Discovery of Very High-Energy Gamma-Ray Emission from RGB J2243+203 }

\ShortTitle{VHE Gamma-Ray Emission from RGB J2243+203}

\author{\speaker{A. U. Abeysekara} for the VERITAS collaboration\thanks{veritas.sao.arizona.edu}\\
	   Department of Physics and Astronomy, University of Utah, Salt Lake City, UT 84112, USA
        E-mail: \email{udaraabeysekara@yahoo.com}}

\abstract{
RGB J2243+203 is a blazar at an estimated redshift of greater than 0.39 that has been classified both as an intermediate-frequency-peaked BL Lac object and as a 
high-frequency-peaked BL Lac object.
This source has been detected by VERITAS at a statistical significance of 5.6 $\sigma$ in 
$\sim$4 hours between 21 Dec 2014 UTC (57012 MJD) and 24 Dec 2014 UTC (57015 MJD). 
An integral flux of $2.03 \pm 0.54 \times 10^{-11} \textrm{cm}^{-2}\textrm{s}^{-1}$ has been observed above 160 GeV.
The gamma-ray spectrum can be described as a simple power-law spectrum with spectral index of $-4.64 \pm 0.56$.
VERITAS measurements are complemented by quasi-simultaneous \textit{Fermi}-LAT and \textit{Swift-}XRT observations, 
in the energy ranges of 1-100 GeV and 2-10 keV, respectively. 
During the three day time period of VERITAS observations, an increased flux level at 1-100 GeV energies was detected compared to the flux from the first four years of \textit{Fermi}-LAT observations. 
The \textit{Swift}-XRT observations were taken 2.8 days after this period, and the integral X-ray flux in the energy range 0.3-10 keV is not significantly higher than a baseline flux from archival data.
}

\FullConference{The 34th International Cosmic Ray Conference,\\
		30 July- 6 August, 2015\\
		The Hague, The Netherlands}

\begin{document}

\section{Introduction}
RGB J2243+203 is a BL Lac type blazar at an estimated redshift of $z>0.39$ \\
\cite{MeisnerAndRomani}.
Blazars are active galactic nuclei (AGNs) with jets emanating from the central black hole pointed towards earth.
BL Lac type objects are a subclass of AGNs that are characterized by there highly variable non-thermal spectra, 
lack of prominent emission lines, and significant optical polarization that varies with time \cite{Kollgaard, UrryAndPadovani}.
Blazars are interesting laboratories for studying AGN phenomena as they are known to have a non-thermal radiation that 
spans all the way from the radio to the gamma-ray regime. 
 
In the literature, BL Lac objects are often classified into categories based on their spectral energy density.
BL Lacs with their low-energy peak in the far IR or IR energy band, typically below $10^{14}$ Hz, are categorized as low-frequency-peaked BL Lacs (LBLs). 
When the low-energy peak is in the range $10^{14} ~ \textrm{Hz} < \nu < 10^{15} ~ \textrm{Hz} $ they are called intermediate-frequency-peaked BL Lacs (IBLs). 
When the low-energy peak is above $10^{15}$ Hz they are called high-frequency-peaked BL Lacs (HBL)  \cite{Spurio2015, Ackermann2013}.
Some other literature, eg. \\ \cite{Laurent1999} and \cite{Padovani1995}, categorized BL Lacs according to the ratio 
of X-ray to radio flux densities.
The boundary between LBL and HBL is drawn at $\log (S_x/S_r) = -5$, 
where $S_x$ is the X-ray flux in the 0.3-3.5 keV energy range in units ergs s$^{-1}$ cm$^{-2}$ and $S_r$ is the 5 GHz radio flux density in Jy.

In 1999, Laurent-Muehleisen et al. studied an extensive sample of BL Lacs, including RGB J2243+203, identified from the 
ROSAT All-Sky Survey-Green Bank (RGB) catalog \\ \cite{Laurent1999}.
The median $\log (S_x/S_r)$ of the RGB BL Lac sample was $-5.61$, which is near the LBL-HBL boundary.
Therefore the RGB BL Lac sample, including RGB J2243+203, were classified as intermediate-frequency-peaked BL Lac objects.

\textit{Fermi}-LAT is a pair-conversion high-energy gamma-ray detector on the \textit{Fermi} gamma-ray space telescope that is
sensitive to gamma-rays in the energy range from 20 MeV to more than 300 GeV, and has a large field of view of 2.4 sr \cite{Atwood2009}.
In 2013, the first \textit{Fermi}-LAT catalog of sources above 10 GeV was published, and the \textit{Fermi}-LAT source 1FHL J2244.0+2020 is associated with RGB J2243+203 \cite{Ackermann2013}.
The 1FHL catalog categorized BL Lacs based on the location of the synchrotron emission peak, which is different than the criteria used in \cite{Laurent1999}, and categorized the source as a HBL object.

\textit{Fermi}-LAT nominally operates in survey mode, where it observes the whole sky in $\sim$3 hr with almost uniform exposure.
The VERITAS (Very Energetic Radiation Imaging Telescope Array System) operates an automated daily analysis of \textit{Fermi}-LAT observations for identifying elevated flux states of sources.
After identifying a marginally elevated flux state, VERITAS started observing RGB J2243+203 on 21 December 2014, UTC.

\section{Observations and Results}
\subsection{VERITAS}

VERITAS is an array of four 12 m imaging atmospheric Cherenkov telescopes located at the Fred Lawrence Whipple 
Observatory (FLWO) in southern Arizona ($31^\circ$ $40^\prime$ N, $110^\circ$ $57^\prime$ W,  1.3 km a.s.l.) \cite{Holder2008}.
VERITAS is sensitive to gamma-rays in the energy range from $\sim$85 GeV to $\sim$30 TeV. 
A point source with brightness of 1\% Crab nebula flux can be detected within 
25 hours with a statistical significance of 5 standard deviations $\left( 5 \sigma \right)$.
The energy of a gamma-ray can be measured with a resolution $15-25\%$.
A source can be localized with an accuracy better than 50 arcsec, 
and the angular resolution, that contains 68\% of the selected events, is better than 0.1 degree.

\begin{figure}[h]
\centering
 \includegraphics[width=0.5\textwidth]{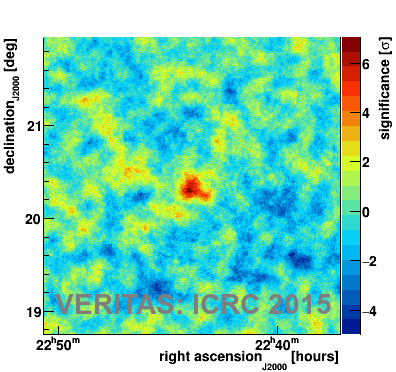}
\caption{Preliminary skymap of the region around RGB J2243+203.}
\label{Fig:VERITASSigmap}
\end{figure}

VERITAS started observing RGB J2243+203 on 21 December 2014 UTC (57012 MJD). 
From the $21^{\textrm{st}}$ (57012 MJD) through the $24^{\textrm{th}}$ (57015 MJD) VERITAS observed the source every night starting from the beginning of the night until the elevation of 
the source went below $40^\circ$.
In the first night, a marginally significant excess (4.2 $\sigma$) was observed from the direction of the source in 37 minutes. 
After observing the source over the next three days for additional 244 minutes, the cumulative significance increased to 5.6 $\sigma$, 
representing the significant detection of the blazar.
We ceased observing  the source on the $24^{\textrm{th}}$, because the moon precluded low-energy-threshold observing.
After the quality cuts the total exposure on the source was 280 minutes live time.
The significance map of the region, centered at the source location is shown in Figure ~\ref{Fig:VERITASSigmap}.
The spectrum of gamma-rays above 160 GeV can be fitted with a simple power-law distribution. 
The best fit has a $\chi^2/NDF$ of 0.3/3, and spectral index of $-4.64 \pm 0.55$.
The integrated gamma-ray photon flux above 160 GeV is $2.03 \pm 0.54 \times 10^{-11} \textrm{cm}^{-2}\textrm{s}^{-1}$.
Detailed light-curves of the VERITAS measurements will be published elsewhere.

Before the 2014 observations, RGB J2243+203 was observed by VERITAS in 2009 
September and October with an exposure of about 5.3 hours.
The past observations showed no evidence for TeV emission from this source.
A 99\% confidence level flux upper limit above 160 GeV of 7.0 $\times 10^{-12} \textrm{cm}^{-2}\textrm{s}^{-1}$ can be calculated from the non-detection.

\subsection{\textit{Fermi}-LAT}
RGB J2243+203 was in the field of view of the \textit{Fermi}-LAT's for about 90 minutes during the 280 minutes that VERITAS observed the source.
In order to get a quasi-simultaneous spectral measurement, we analyzed the \textit{Fermi}-LAT data set starting from 
2014-12-21 01:44:00 UTC to 2014-12-24 03:34:00 UTC (57012.07222222 MJD to 57015.14861111 MJD), overlapping the time of the VERITAS observation of RGB J2243+203.
Photons between energy 1 GeV and 100 GeV that belong to \textit{Fermi}-LAT event class 2 within a region of interest (ROI) 
$10^\circ \times 10^\circ$ centered at RGB J2243+203 were selected.
Selection cuts of rocking angle $< 52^\circ$, zenith angle $< 100^\circ$, minimum photon energy of 1 GeV, and maximum 
photon energy of 100 GeV were used to filter data.
The data set has been analyzed with publicly available \textit{Fermi}-LAT analysis tools with (v9r33p0) standard instrument response
function P7REP\_SOURCE\_V15.
The galactic diffuse emission and the extragalactic isotropic diffuse emission was modeled with gll\_iem\_v05\_rev1.fit and 
so\_source\_v05.txt, respectively.
Modeling of the gamma-ray sources in the ROI is started with the \textit{Fermi} 2FGL catalog sources in the ROI.
After the first iteration all the point sources with TS $<$ 0 are removed from the model, and the likelihood analysis is re-performed.
The best-fit \textit{Fermi}-LAT results converged to an integral flux between 
1 GeV and 100 GeV of $2.4 \pm 1.0 \times 10^{-8} \textrm{cm}^{-2} \textrm{s}^{-1}$, and photon index of $-2.9 \pm 0.7$, 
with a test statistic of TS = 22.2.

\subsection{\textit{Swift}-XRT}
The focusing X-ray telescope (XRT) on board \textit{Swift} \cite{Gehrels2004} is sensitive to 0.3 to 10 keV photons.  
Triggered by the VERITAS detection of the blazar, \textit{Swift} observed RGB J2243+203 five times between Modified Julian Date (MJD) 57017.945 and 
57021.735.
The observed count rates ranged between 0.03 and 0.08 counts per second, and are not significantly different. 
Four archival observations, taken between MJD 54956.2 and 56842.811 were also analyzed.
The observed count rates ranged between 0.02 and 0.3 counts per second.
All observations were carried out using the Photon Counting (PC) readout mode, and did not suffer from any photon pileup.  
The data were calibrated and cleaned with standard criteria with the {\it xrtpipeline} task using the calibration 
files as available in the \textit{Swift} CALDB version 20140120.

The observation specific spectra were extracted from the summed and cleaned event files.  
Events for the spectral analysis were selected within a circle of 20 pixels, corresponding to $\sim 45 ''$ radius, 
centered on the source position. 
The background was extracted from a nearby circular region of 40 pixel radius.  
The most recent response matrices (v.014) available in the \textit{Swift} CALDB were used. 
Before spectral fitting of the summed observation file, for exposures between MJD 57017.945 and 57021.735, 
the spectra were binned to require a minimum of 20 counts per bin, allowing for $\chi^2$ minimization 
fitting procedure in spectral fitting.
The data were fit with an absorbed power-law model, with index $\Gamma$, with a neutral hydrogen column density set to 
$5.16 \times 10^{20}$cm$^{-2}$, from \cite{kalberla}.
The best fit spectrum has a $\Gamma = -2.7 \pm 0.2$, and the integrated photon flux between 2 keV and 10 keV are 
$3.58 \pm 0.6 \times 10^{-13} \textrm{cm}^{-2} \textrm{s}^{-1}$. 
In the four archival observations, photon fluxes between 2 keV and 10 keV of  
$26.3 \pm 1.4 $, 
$4.3  \pm 1.1$, 
$3.5  \pm 2$, and
$2.7  \pm 0.7 \times 10^{-13} \textrm{cm}^{-2} \textrm{s}^{-1}$, respectively, were measured.

\section{Discussion}
The multi-wavelength measurements of the source are summarized in Table~\ref{MultiWaveLengthProp}.
The high-energy gamma-ray photon flux of RGB J2243+203 detected by VERITAS, above 160 GeV, during the high-flux state was 
factor of 3 larger than the upper limit placed by VERITAS for the observations taken before 2014. 
In the \textit{Fermi}-LAT third source catalog \cite{FermiLAT3FGL}, the source was labeled as 3FGL J2243.9+2021.
The average flux between 1 and 100 GeV in the \textit{Fermi}-LAT 4 year data set is $4.0 \pm 0.1 \times 10^{-9} 
\textrm{cm}^{-2} \textrm{s}^{-1}$ with a variability index is 59.1.
A variability index less than 72.44 suggest that the source was mostly steady \cite{FermiLAT3FGL}.
During period of the VERITAS observations, the measured flux in the 1-100 GeV energy range is factor of six larger than 
the average flux.
These GeV measurements suggest that the source was in an active state during the VERITAS observations.
Detailed analysis of the \textit{Fermi} LAT variability and the light curve will be published elsewhere.

\begin{figure}
\centering
\includegraphics[width=0.8\textwidth]{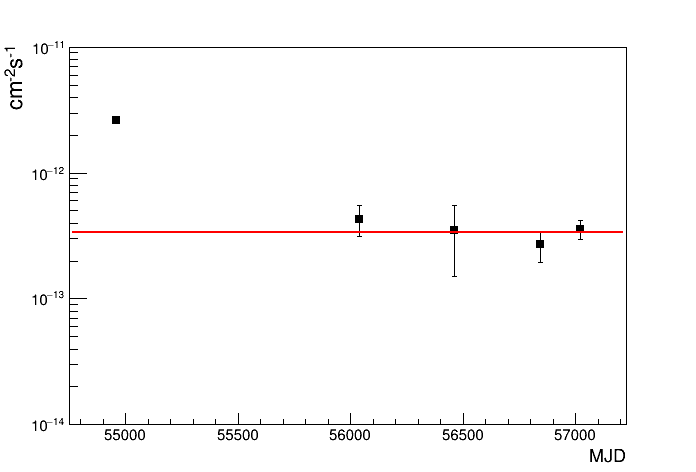}
\caption{The photon flux between 2 and 10 keV measured by \textit{Swift}-XRT of the source at four different times is shown. The y-axis is the integrated flux between 2 and 10 keV. The x-axis is the MJD of the start time of the observation. 
The last data point is the quasi-simultaneous measurement to the VERITAS observation of the source.}
\label{Fig:XRT}
\end{figure}

Figure \ref{Fig:XRT} shows the \textit{Swift}-XRT soft X-ray measurements of the source, 
including four flux measurements taken before the VERITAS observation, and the flux measured quasi-simultaneously to the VERITAS observation of the source.
These five flux measurements were analyzed independently.
The red line is the best fit of a constant flux to the last four data points.
The first data point is not consistent with a constant flux.
However, the last four data points are consistent with a constant flux, including the data point quasi-simultaneous to the VERITAS observations of the source.
Therefore, we conclude that the X-ray flux, quasi-simultaneous to the VERITAS observations, in the energy range 1-10 keV is not significantly higher than the baseline flux from the archival data.
One has to note that \textit{Swift}-XRT observed the source about 67 hours after VERITAS and \textit{Fermi}-LAT observed the elevated state.
Therefore, it is possible that \textit{Swift}-XRT missed any X-ray activity the source may have displayed.

\begin{table}[h]
\centering
\begin{tabular}{ |l | c | c | c | }
  \hline
  Instrument & Energy range & Flux $\left( \textrm{cm}^{-2} \textrm{s}^{-1} \right)$ & Power-law index\\ \hline
  VERITAS & E > 160 GeV &  $2.03 \pm 0.54 \times 10^{-11}$ & $-4.64 \pm 0.56$ \\
  \textit{Fermi}-LAT & 1 - 100 GeV & $2.4 \pm 1 \times 10^{-08}$ & $-2.9 \pm 0.7$ \\
  \textit{Swift}-XRT & 2 - 10 keV & $3.58 \pm^{2.9}_{0.6} \times 10^{-13}$ & $-2.7 \pm 0.2$ \\
  \hline  
\end{tabular}
\caption{Quasi-simultaneous multi-wavelength measurements of RGB J2243+203. \label{MultiWaveLengthProp}}
\end{table}

\section{Conclusion}
The observations of RGB J2243+203 with VERITAS in 2014 December resulted in the discovery of very high energy gamma-rays from the source.
Multi-wavelength observations quasi-simultaneous to VERITAS observations were collected from \textit{Fermi}-LAT and \textit{Swift}-XRT.
VERITAS and \textit{Fermi}-LAT data shows that the source was in an active gamma-ray state during the observations.
There was no X-ray activity observed by \textit{Swift}-XRT.
It is possible that \textit{Swift}-XRT missed the active state of the source due to the delay in XRT exposures as compared to the elevated gamma-ray detection.

\section{Acknowledgements}
This research is supported by grants from the U.S. Department of Energy Office of Science, 
the U.S. National Science Foundation and the Smithsonian Institution, 
and by NSERC in Canada. 
We acknowledge the excellent work of the technical support staff at the Fred Lawrence Whipple Observatory and at the 
collaborating institutions in the construction and operation of the instrument.
The VERITAS Collaboration is grateful to Trevor Weekes for his seminal contributions 
and leadership in the field of VHE gamma-ray astrophysics, which made this study possible.

\end{document}